\documentclass[lettersize,journal]{IEEEtran}

\usepackage[ left=0.625in, top=0.75in, right=0.75in, bottom=1.25in, headsep=0.25in, headheight=20pt, heightrounded, letterpaper] {geometry}
\usepackage[utf8]{inputenc}
\usepackage{lipsum}
\usepackage{comment}
\usepackage{xfrac}  

\usepackage{lipsum,microtype}
\usepackage{authblk}
\usepackage[ruled,vlined]{algorithm2e}
\usepackage{amsmath, nccmath}
\usepackage{geometry}
\usepackage{tabularx}
\usepackage{booktabs}
\usepackage{setspace}
\usepackage[labelfont=bf,format=plain,justification=raggedright,singlelinecheck=true]{caption}
\usepackage{mathtools}
\usepackage{amssymb}
\usepackage{caption}
\usepackage{xcolor}
\usepackage{subcaption}
\usepackage{graphicx}
\graphicspath{{./Fig/}} 
\newcommand*{\Scale}[2][4]{\scalebox{#1}{$#2$}}%

\usepackage[ruled,vlined]{algorithm2e}
\usepackage{cite}
\newcounter{mycounter}
\usepackage{subfig} 
\usepackage{fancyhdr}
\usepackage{textcomp}
\usepackage{blindtext}
\usepackage{optidef}
\usepackage{subcaption} 
\usepackage{ragged2e} 
\usepackage{multirow}
\usepackage{longtable}
\usepackage{pifont}

\usepackage{fancyhdr}
\usepackage{textcomp}
\usepackage{blindtext}

\title{Smart Jamming Attack and Mitigation on Deep Transfer Reinforcement Learning Enabled Resource Allocation for Network Slicing}

\author{Shavbo~Salehi,~\IEEEmembership{ Student Member, IEEE},
        Hao~Zhou,
        Medhat~Elsayed,
        Majid~Bavand,
        Raimundas~Gaigalas,
        Yigit~Ozcan,
        and~Melike~Erol-Kantarci,~\IEEEmembership{Senior Member, IEEE}
\thanks{Shavbo Salehi, Hao Zhou, and Melike Erol-Kantarci are with the School of Electrical Engineering and Computer Science, University of Ottawa, Ottawa, ON K1N 6N5, Canada (e-mail: ssale038@uottawa.ca; hzhou098@uottawa.ca; melike.erolkantarci@uottawa.ca).}
\thanks{Medhat Elsayed, Majid Bavand, and Yigit Ozcan are with Ericsson, Ottawa, K2K 2V6, Canada(e-mail:medhat.elsayed@ericsson.com; majid.bavand@ericsson.com; yigit.ozcan@ericsson.com) 
}
\thanks{Raimundas Gaigalas, Ericsson AB, Stockhom, Sweden(e-mail: raimundas. gaigalas@ericsson.com) 
}}     

\date{}

\begin{document} 

\maketitle

\begin{abstract}

Network slicing is a pivotal paradigm in wireless networks enabling customized services to users and applications. Yet, intelligent jamming attacks threaten the performance of network slicing. In this paper, we focus on the security aspect of network slicing over a deep transfer reinforcement learning (DTRL) enabled scenario. We first demonstrate how a deep reinforcement learning (DRL)-enabled jamming attack exposes potential risks. In particular, the attacker can intelligently jam resource blocks (RBs) reserved for slices by monitoring transmission signals and perturbing the assigned resources. Then, we propose a DRL-driven mitigation model to mitigate the intelligent attacker. Specifically, the defense mechanism generates interference on unallocated RBs where another antenna is used for transmitting powerful signals. This causes the jammer to consider these RBs as allocated RBs and generate interference for those instead of the allocated RBs. The analysis revealed that the intelligent DRL-enabled jamming attack caused a significant 50\% degradation in network throughput and 60\% increase in latency in comparison with the no-attack scenario. However, with the implemented mitigation measures, we observed 80\% improvement in network throughput and 70\% reduction in latency in comparison to the under-attack scenario. 
\end{abstract}

\begin{IEEEkeywords}
network slicing, deep transfer reinforcement learning, intelligent jamming attack, and jamming attack mitigation.    \vspace{-0.3cm}

\end{IEEEkeywords}

\begin{table}[ht]
    \centering
    \caption{List of Abbreviations}
    \label{tab:abbreviations}
    \small
    \begin{tabular}{|l|p{6cm}|}
    \hline
    \textbf{Abbreviation} & \textbf{Elaboration} \\
    \hline
    DTRL & Deep Transfer Reinforcement Learning \\
    RB & Resource Block \\
    eMBB & enhanced Mobile Broadband  \\
    uRLLC & ultra-Reliable Low-Latency Communication  \\
    mMTC & massive Machine-Type Communication \\
    gNB & Next Generation Node Base  \\
    MDP & Markov Decision Process \\
    LSTM & Long Short-Term Memory \\
    DQN & Deep Q-Network  \\
    RSS & Random State Selection  \\
    RPQL & Reward Poisoning against Q-learning  \\
    BRPQL & Bounded Reward Poisoning against Q-learning \\
    CPJA & Covert Patterned Jamming Attack \\
    UE & User Equipment\\
    MEC & Mobile Edge Computing  \\
    CJA & Constant Jamming Attack  \\
    RJA & Random Jamming Attack  \\
    DRL-JA & DRL-based adaptive Jamming Attack  \\
    eCDF & empirical Cumulative Distribution Function \\
    HetNets & Heterogeneous networks \\
    UL and DL & Uplink and Downlink \\
    \hline
    \end{tabular}
\end{table}

\begin{table*}[ht]
    \centering
    \caption{MODEL PARAMETERS}
    \label{tab:notations}
    \small
    \begin{tabular}{|l|p{5.8cm}|l|p{5.5cm}|}
    \hline
    \textbf{Parameters} & \textbf{Explanation} & \textbf{Parameters} & \textbf{Explanation} \\
    \hline
    $w^{eMBB}$ & Weighting factor eMBB & $w^{uRLLC}$ & Weighting factor uRLLC \\
    $B^{eMBB,avg}_{j}$ & Average throughput of eMBB slice in gNB $j$ & $D^{uRLLC,avg}_{j}$ & Average delay at uRLLC slice in gNB $j$ \\
    $q^{eMBB}$ & eMBB requests in queue & $q^{uRLLC}$ & uRLLC request in queue \\
    $r^{eMBB}$ & Assign RBs to eMBB UEs & $r^{uRLLC}$ & Assign RBs to uRLLC UEs \\
    $D^{tar}$ & Target delay of uRLLC slice & $(s_t, a_t, r_{t+1}, s_{t+1})$ & (State, Action, Reward, Next state) \\
    $\alpha$ & Learning rate & $\gamma$ & Discount factor \\
    $s_{l,t}$ & Current state learner agent at time $t$ & $a_{l,t}$ & Current action learner agent at time $t$\\
    $F$ & Mapping functions applied to states & $F'$ & Mapping functions applied to actions \\
    $\lVert RBs\lVert$ & The length of RBs & $T_{KB}$ & Knowledge repository \\
    $p_{j,r}$ & Transmission power of RB $r$ in gNB $j$ & $Q^e(s_e, a_e)$ & Q-value, states, and actions of expert agent \\
    $x_{j,u,r}$ & Designated RB $r$ for UE $u$ & $T^{th}$ $gNB$ & Targeted gNB for CJA \\
    $N_{0}$ & Density of noise power & $N_{j'}$ & Entire RBs in gNB $j'$ \\
    $P_{\psi_{CJA}}$ & Transmit power of CJA signal  &  $T_{sl}$ & Sensing slot \\
    $E_{\theta}$ & Received power & $P_{T^{th}}$ & Transmission power of the $T^{th}$ \\
    $M_{DRL-JA}$ & MDP of DRL-JA & $s_{DRL-JA}$ & State of DRL-JA \\
    $a_{DRL-JA}$ & Action of DRL-JA & $r_{DRL-JA}$ & Reward of DRL-JA \\
    $X_r$ & Maximum energy detection threshold & $b_q$ & Bandwidth of RB $q$ \\
    $\widetilde{P}_r$ & Exerted interference power of DRL-JA & \(\omega_1\) &  Prioritize interference control weight\\
    \(\widetilde{L}\) & System \(SINR\) after the DRL-JA &  \(\omega_2 \widetilde{E}\) & Weight of energy consumption metrics \\
    $T_{DRL-JA}$ & Duration of the DRL-JA transmission & $\pi$ & Policy function \\
    $\gamma_{DRL-JA}$ & Discount factor of the DRL-JA & $\pi_{DRL-JA}^*$ & Optimal policy of the DRL-JA \\
    $s_m$ & Mitigation method's states & $a_m$ & Mitigation method's action set \\
    $r_m$ & Mitigation method's reward & $\mathbb{I}$ & Learning rate of mitigation method\\
    $\pi_{m}^*$ & MDP of the security administrator & $\chi$ & Deep Q-learning \\
    $P_{T^{th}}$ & Transmission power of the no-attack system & $L{{j,u}}$ & Interference power of the no-attack system \\ 
    $P_{DRL_{mit}}$ & Antenna power consumption &$p_i$ & Power for interference signal generation \\ 
    $\mathfrak{B}$ & Targeted RBs for attack & $\xi[n]$ & Power of the discrete signal \\ 
    \hline
    \end{tabular}
\end{table*}

\section{Introduction}
\label{section:intro}

The deployment of 5G and beyond 5G technology has ushered in an era of connectivity, promising unprecedented levels of service customization, efficiency, and user experience. One of the key innovations in this transformation is network slicing \cite{filali2022dynamic}, which creates logically independent network segments within a shared infrastructure \cite{bonati2020open}. Network slicing can be used for enhanced mobile broadband (eMBB), ultra-reliable low-latency communications (URLLC), massive machine-type communications (mMTC) \cite{liu2023network}, or other new service types. In network slicing, resources are allocated efficiently to ensure the effectiveness of URLLC, eMBB, and MTC applications \cite{jankovic2021effects}. A promising solution for resource allocation lies in machine learning (ML), which provides a dynamic and adaptable approach \cite{elsayed2019ai}. By learning patterns from past data, predicting future resource demands, and adapting to changing network conditions, ML algorithms play an integral role in optimizing resource slicing \cite{sharma2019toward}. Particularly reinforcement learning (RL) and deep learning (DL) are of particular importance due to their ability to learn patterns and exceptionally adept to dynamic network conditions and optimizing resource allocation \cite{hussain2020machine}. Above this, the power of ML emerges even more prominently through transfer learning (TL) and deep transfer reinforcement learning (DTRL)\cite{zhou2022learning}. TL and DTRL allow models to use knowledge learned from one agent to enhance the performance of learning in other agents.

It is undeniable that ML algorithms improve the performance of wireless networks however both wireless systems and ML algorithms are vulnerable to adversarial challenges, particularly jamming attacks. Jamming attacks can be classified as constant, random, deceptive, or reactive, each with its methodology and implications for network security \cite{pirayesh2022jamming}. Note that, jamming attacks are characterized by deliberate interference with wireless communications and they pose a substantial threat to the communication. There has been some research conducted on the effect of jamming attacks on ML-based network slicing systems. Based on RL-based network optimization, the authors in \cite{shi2022jamming} examine a jamming attack on resource blocks (RBs) by focusing on reducing the RL algorithm’s reward in network slicing. In \cite{shi2021attack}, the jamming attack on an RL-enabled network slicing scenario is considered in which the attacker, by jamming RBs, tries to maximize the number of failed requests. Although the vulnerability of RL-based approaches has been studied, to the best of our knowledge vulnerabilities of DTRL to jamming attacks have not been explored. 
In the context of DTRL-based network slicing, such disruptions can result in sub-optimal decision-making, inefficient resource allocation, and compromised service quality. In addition, these attacks and malicious knowledge can be transferred between systems, where such knowledge transfer between experts and learners expands the attack surface. Furthermore, these attacks may become more impactful if they are also designed with some intelligence, for instance using ML techniques. In this paper, 
we examine the effect of the DRL-based jamming attack while the DTRL system provides services to eMBB and URLLC slices. By monitoring the transmission signal of the targeted next generation node bases (gNB), the attacker can distinguish the RBs allocation. Then the attacker transmits a malicious interference signal, which degrades the normal channel quality. The DRL-based jamming attacks exhibit distinct characteristics that negatively impact the functionality and reliability of the network. The intelligent attack involves the deliberate transmission of interference signals on the same frequency bands utilized by legitimate network communications. Due to using DTRL for slicing decisions, the malicious knowledge by the jamming attack is transferred from the expert to the learner agent, which means the attack spreads between intelligent agents. Therefore, mitigating intelligent jamming attacks is challenging and important to ensure the robustness and security of the system.

DRL-based jamming attacks are dynamic and diverse \cite{wang2023sensitivity, wang2022robust}, requiring real-time responses. Consequently, the need for ML-based mitigation methods arises, ensuring intelligent and fast responses to effectively counter these attacks. In \cite{ismail2021jamming} a prioritized double-deep Q-learning defense strategy is proposed against jamming attacks by learning the jammer's capacity and attack timing. This strategy involves learning the jammer's capacity and attack timing. However, both the federated DRL and prioritized double-deep Q-learning methods require distributed data for a comprehensive system observation. Moreover, previously proposed methods continue to struggle with the issue of missed detection \cite{benslimane2016jamming}. Missed detections occur when the system fails to identify a genuine attack highlighting the importance of developing mitigation methods to decrease the probability of such occurrences. As a means of mitigating intelligent attacks, in this paper, a DRL-based model is designed to mislead the attacker to generate interference signals on vacant RBs and increase its miss detection rate on assigned RBs. Enhanced by DRL, our method is implemented in a gNB, which gathers essential information for robust attack mitigation techniques. Meanwhile, our DRL-based technique misleads the attacker to transmit an interference signal to unallocated RBs. This technique relies on deception to induce attackers toward unproductive actions, thereby enhancing the overall defense posture, whereas traditional mitigation strategies are focused on detection and prevention. The main contributions of this paper are as follows:

\begin{itemize}
    \item Novel Intelligent Jamming Attack (DRL-JA): Introduces a dynamic DRL-based jammer that strategically monitors and targets gNB's allocated RBs. This intelligent attack, unlike random or constant jammers, adapts its policy in real-time, significantly intensifying disruption in DTRL systems.
    \item Advanced DRL-Based Mitigation: Proposes a robust mitigation method that leverages DRL to generate misleading signals on unallocated RBs, deceiving the DRL-JA. This dynamic defense strategy preserves system performance by reducing the attack's impact on legitimate RB allocations.
    \item LSTM-Powered Predictive Defense: Incorporates an LSTM network within the mitigation framework, enabling the security administrator to anticipate jammer behavior and adjust the mitigation strategy accordingly, ensuring timely and adaptive responses to attacks.
    \item Comprehensive Impact Analysis: Conducts an in-depth evaluation of DRL-JA's effects on a DTRL-based system, demonstrating how the attack undermines system efficiency, particularly affecting the learner agent’s resource allocation capabilities.
    \item Simulation and Performance Recovery: Provides extensive simulations illustrating that the proposed DRL-based mitigation with LSTM curbs interference, restores system performance, and outperforms traditional mitigation approaches in complex jamming scenarios.
\end{itemize}

\section{Related Work}
\label{section:LR}

\begin{table*}
    \centering
\caption{Comparison of Jamming Attack Papers}
\label{tab:Methods}
\small
\setstretch{1.1}
\resizebox{1\textwidth}{!}{
    \begin{tabular}{|>{\centering\arraybackslash}m{0.8cm}|m{2.6cm}|m{1.8cm}|m{4.4cm}|m{3cm}|m{4.4cm}|}
\hline
\textbf{Ref} & \textbf{System \newline Scenario} & \textbf{Attack Mechanism} & \textbf{Strengths \& Weaknesses} & \textbf{Mitigation Techniques} & \textbf{Strengths \& Weaknesses}\\
\hline
\cite{di2013jamming} & CR Networks & - & - & Anti-jamming for CRNs & \checkmark Independent channel information \newline \ding{53} Propagation delay \\ 
\hline
\cite{jover2013security} & Cellular \newline Networks & DoS, DDoS & \checkmark Ease of execution,\newline \ding{53} Limited range, easy detection & - & - \\ 
\hline
\cite{grover2014jamming} & Ad-hoc \newline Networks & Constant \newline jamming & \checkmark Highly effective, \newline \ding{53} Energy inefficient & - & - \\ 
\hline
\cite{shahriar2014phy} & OFDM \newline Networks & Preamble jamming & \checkmark Stealthy, energy efficient \newline \ding{53} Tight timing, hard implement & - & -\\ 
\hline
\cite{devi2014denial} & Ad-hoc \newline Networks & Constant \newline jamming & \checkmark Highly effective, \newline \ding{53} Energy inefficient & - & -\\ 
\hline
\cite{lichtman20185g} & 5G New \newline Radio & Jamming, spoofing & \checkmark Impairs performance, \newline \ding{53} Limited affects, detectable & Frequency-hopping & \checkmark Shift communication \newline \ding{53} Complexity, additional resource\\ 
\hline
\cite{hachimi2020multi} & Cloud Radio Networks & - & - & Multi-stage ML detection &  \checkmark High accuracy \newline \ding{53} False negative, data dependency\\ 
\hline
\cite{arjoune2020novel} & Wireless \newline Networks & - & - & SVM, NN, random forest & \checkmark Low false alarm rate \newline \ding{53} Data dependence, smart jamming inefficient\\ 
\hline
\cite{shi2018adversarial} & CRN & Adversarial DL, random & \checkmark Adaptive \newline \ding{53} Resource and observation dependent & Taking incorrect actions & \checkmark Sustain performance \newline \ding{53} Scarifies accuracy\\ 
\hline
\cite{erpek2018deep} & Wireless \newline Networks & GAN-based jamming & \checkmark Energy efficient \newline \ding{53} Complex, data-dependent & Noise injection & \checkmark Jammer disruption \newline \ding{53} Throughput reduction \\ 
\hline
\cite{sharma2022mitigating} & 5G HetNet & - & - & Federated DRL-based & \checkmark Self-adaptive capabilities \newline \ding{53} Resource consumption, slow convergence, complexity\\ 
\hline
\cite{shi2022launch} & Wireless \newline Networks & UL/DL \newline jamming & \checkmark Scalable, minimal data, dynamic \newline \ding{53} Energy consumption, FL resilience, limit budget & - & - \\ 
\hline
\cite{shi2022jamming} & Wireless \newline Networks & UL/DL \newline jamming & \checkmark Versatile \newline \ding{53} Limit budget, low effect & - & - \\ 
\hline
\cite{shi2023jamming} & Wireless \newline Networks & MCBA, NDBA, random & \checkmark Effective, scalable \newline \ding{53} Resource-limited, multi-hop resilience & -& - \\ 
\hline
\cite{shi2022jamming2} & Network \newline Slicing & Q-learning, random, myopic & \checkmark Prolonged impact \newline \ding{53} Energy budget constraints, low effect & - & -\\ 
\hline
\cite{ICC2024} & Network\newline Slicing & CPJA, constant jamming & \checkmark Disruptive, covert \newline \ding{53} Specific impact & NN-based mitigation & \checkmark Predictive capability, deployment flexibility \newline \ding{53} Training and adaptation\\ 
\hline
\textbf{[Our work]} & DTRL Network\newline Slicing & DRL-JA, random, constant jamming & \checkmark Effective, scalable, detection-resistant, environment awareness, adaptive strategy  \newline \ding{53} DRL model dependent & DRL-based mitigation, learning suspending & \checkmark Effective, performance preserve, intelligent adaptation\newline \ding{53} Resource budget constraints\\ 
\hline
\end{tabular}}
\end{table*}

ML has proven to be an effective tool for improving various aspects of network slicing, especially those based on RL and DL\cite{yan2019intelligent}. These algorithms optimize resource allocation, manage dynamic network conditions, and satisfy user requirements \cite{thantharate2023adaptive6g}. By utilizing data and learning patterns, network slicing systems powered by TL improve network efficiency, reduce latency, and enhance the overall experience for users, according to \cite{hu2023inter}. In addition, TL-based algorithms enable gNBs to make intelligent decisions, allocate resources dynamically, and respond to network traffic fluctuations. Besides, DTRL algorithms have emerged as a promising solution for improving resource allocation and management in network slicing \cite{zhou2022learning}. 

Although network slicing based on ML has many advantages, it is also prone to security vulnerabilities \cite{de2023survey}. Note that, the vulnerabilities are not a result of slicing but may increase due to the implementation of network slicing using ML techniques \cite{vila2022impact}. There has been some research conducted on the effect of jamming attacks on ML-based network slicing systems. Jamming attacks, in particular, can be commonplace within this context, disrupting wireless communication channels intentionally \cite{lu2021secure}. These attacks disrupt or interfere with wireless signals to cause communication breakdowns, service interruptions, and compromised network functionality \cite{sharma2022mitigating}. As a result, network slicing algorithms allocate erroneous resources to traffic, degrading the network performance and the service quality \cite{shi2022jamming2}. In \cite{wang2022robust}, the vulnerabilities of three DRL-based network slicing frameworks—actor-critic, deep Q-Network (DQN), and soft DQN are examined against random, actor-critic, and DQN attacks. In \cite{shi2021attack}, the jamming attack on the RL-enabled network slicing scenario is considered in which the attacker, by jamming RBs, tries to maximize the number of failed requests. Adversarial attacks are also investigated in our previous work \cite{salehi2023adverserial} and \cite{salehi2024jamming}. In particular, we discussed the effects of three adversarial attacks including random state selection (RSS), reward poisoning against Q-learning (RPQL), and bounded reward poisoning against Q-learning (BRPQL) on a TRL-based network slicing system \cite{salehi2023adverserial}. In \cite{salehi2024jamming}, we considered the effects of the covert patterned jamming attack (CPJA) on the TRL-based network slicing system. However, this paper is different from previous works by 1) considering DRL-enabled jamming attack on DTRL network slicing; 2) DRL-driven mitigation model to mitigate the intelligent jamming attack. To the best of our knowledge, the impact of smart jamming attacks on DRL-based systems has not been previously considered. Compared with RPQL, BRPQL, and CPJA techniques \cite{salehi2023adverserial,salehi2024jamming}, the results show approximately a twofold increase in latency for uRLLC traffic and a halving of throughput for eMBB traffic due to the DRL jamming attack. In the context of DTRL-based network slicing, disruptions result in sub-optimal decision-making, inefficient resource allocation, and compromised service quality. In addition, these attacks and poisonous knowledge can be transferred among systems where knowledge transfer between experts and learners expands the attack surface. Furthermore, these attacks become more damaging if they are also designed with some intelligence, for instance using ML techniques.

To maintain the reliability and resilience of network slicing systems under jamming attacks, it is essential to develop robust security mechanisms and adaptive strategies. There is prior research on mitigation techniques based on DRL, aiming to adapt to evolving jamming attacks in a distributed manner \cite{sharma2022mitigating2,ismail2021jamming}. In \cite{sharma2022mitigating2} a federated DRL approach is proposed for mitigating jamming attacks in a joint optimization problem at femtocell base stations. In \cite{shi2021attack}, the suspension of RL algorithm updates upon detection of an attack, the addition of randomness to decision-making processes, and the manipulation of feedback mechanisms have been explored to mitigate the impact of jamming attacks for network slicing. In \cite{jamming2020}, three methods by harnessing ML algorithms, random forest, support vector machine (SVM), and neural network signal classification, have been investigated to learn patterns of legitimate and jammed signals for detecting attacks. In \cite{salehi2024jamming}, a neural network mitigation solution is proposed for the CPJA in a network slicing system. While various research studies have been made about the vulnerability of network slicing systems \cite{adesina2022adversarial}, there is still a lack of consideration of these attacks on DTRL systems. For the first time, our paper investigates the impact of intelligent jamming attacks on DTRL scenarios. Furthermore, different than the studies that propose DRL-based mitigation techniques against jamming attacks, we propose a novel DRL-based technique that misleads the attacker to transmit an interference signal on unallocated RBs. The reason for this approach stems from the recognition that traditional mitigation strategies often focus on detection and prevention, whereas our technique considers deception, steering the attacker toward unproductive actions and thereby enhancing the overall defense posture. Table. \ref{tab:Methods} listed system baselines, jamming attacks, and mitigation techniques of recent studies. As outlined in this table, the jamming attack problem has been considered across various domains, with different algorithms employed for designing jamming attacks and mitigation. However, among the many studies, only a few have proposed methods for both jamming attacks and their mitigation. In this work, we present the first investigation into the impact of an intelligent jamming attack on a DTRL network slicing scenario. The jammer, in designing the attack, utilizes a DRL-based method, while the security administrator also employs a DRL model for attack mitigation. Using DRL-based methods significantly impacts the expert agent's knowledge, with those effects being transferred to the learner agent. However, on the mitigation side, the DRL-based method enables the system to mislead the jammer and recover system performance.

\begin{figure*}[t!]
\centering
\includegraphics[width=1\linewidth]{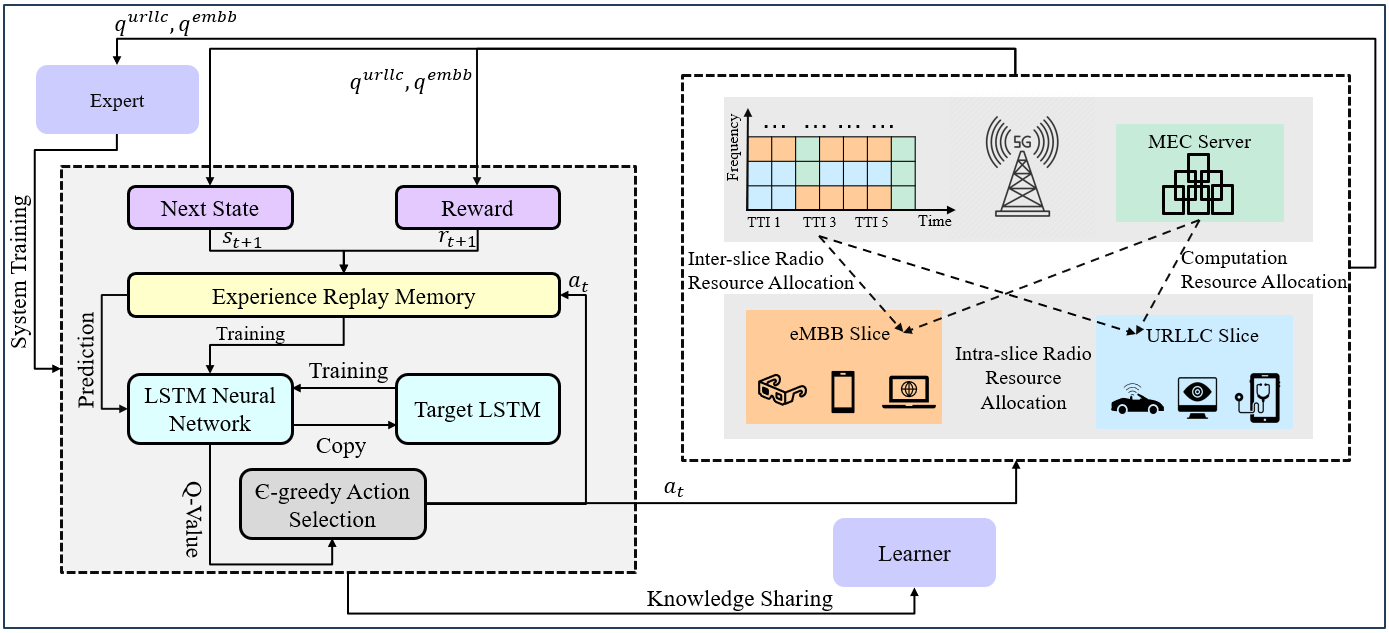}
\caption{Our proposed Network Slicing Scenario.} 
\label{fig:Sce}
\vspace{-0.3cm}
\end{figure*}

\section{System Model for Network Slicing }
\label{section:Scenario}

Before introducing our intelligent attack and mitigation technique, we first provide the system model for the DTRL-based joint radio and computation resource allocation in 5G networks which was proposed in \cite{zhou2022knowledge}. The use of DTRL facilitates efficient and effective resource utilization, while simultaneously accommodating diverse service requirements of eMBBs and uRLLCs slices. The objective of the DTRL model is to facilitate resource allocation of slices through the transfer of expertise from the expert to the learner. 

In our network slicing system model, we assume two types of users and corresponding slices \cite{zhou2022knowledge}. This scenario involves allocating radio and computation resources between two slices to meet the latency and throughput requirements of uRLLC and eMBB user equipment (UE). Radio resources are defined as time-frequency RBs, while computation resources are defined as the available computational capacity used by edge computing. As shown in Fig. \ref{fig:Sce}, gNBs are equipped with mobile edge computing (MEC) servers capable of offloading tasks to the cloud. A channel consists of a frequency band or a portion thereof, including a set of RBs utilized for executing a channel access procedure within a shared spectrum. The significance of spectrum utilization becomes particularly evident when assessing system latency for uRLLC slices which involves considering the following factors:
\begin{equation}
      d = d^{Tx}+d^{rTx}+d^{que}+\eta d^{edge} + (1-\eta)d^{Cloud},
      \label{eqn:delay_total}
\end{equation}
where $d^{Tx}$ denotes the transmission delay and $d^{rTx}$ represents the re-transmission delay. $d^{que}$ signifying the queuing delay, $d^{edge}$ indicating the processing delay within the MEC server, and $d^{Cloud}$ accounting for the processing delay within the central cloud. The binary variable $\eta$ is employed to depict whether the task processing occurs in the MEC server or the cloud. Additionally, $d^{Cloud}$ is affected by the downlink and uplink delay, and the cloud queuing delay, which refers to the amount of time it takes for cloud services to process data. In terms of computation delay, it refers to the time taken by the user's device to process the data.

To meet the requirements of UEs, the uRLLC slice is designed to reduce latency, and the eMBB slice is designed to increase throughput. Based on this consideration, the allocation of RBs formulated as follows: 

\begin{equation}
\Scale[0.95]{
\begin{aligned}
& \max \quad w^{eMBB}B^{eMBB,avg}_{j} + w^{uRLLC}(D^{tar}-D^{uRLLC,avg}_{j}), 
\label{eqn:reward}
\end{aligned}
}
\end{equation}

\begin{equation}
\tag{\arabic{mycounter}.a}
\sum_{u\in M^{eMBB}_{j}} x_{j,u,r'} + \sum_{v\in M^{uRLLC}_{j}} x_{j,v,r'} = 1,
\label{eqn:2a}
\end{equation}

\begin{equation}
\tag{\arabic{mycounter}.b}
\sum_{r' \in N_{j'}} ( \sum_{u\in M^{eMBB}_{j}} x_{j,u,r'} + \sum_{v\in M^{uRLLC}_{j}} x_{j,v,r'} ) \leq |N_{j}| ,
\label{eqn:2b}
\end{equation}

\begin{equation}
\tag{\arabic{mycounter}.c}
C^{eMBB}_{j} + C^{uRLLC}_{j} \leq C_{j},
\label{eqn:2c}
\end{equation}
in Equation (\ref{eqn:reward}), $w^{eMBB}$ and $w^{uRLLC}$ represent the weighting factors assigned to the eMBB and uRLLC slices, respectively, facilitating the comparison of $B^{eMBB,avg}_{j}$ and $D^{uRLLC,avg}_{j}$. The weight of 1 is chosen for both \(w^{eMBB}\) and \(w^{uRLLC}\) factors to balance their influence, reflecting their comparable scales in average uRLLC latency (around 1 \(\text{ms}\)) and eMBB throughput (around 1 \(\text{Mbps}\)) observed in the simulation results. Two considered variables $B^{eMBB,avg}_{j}$ and $D^{uRLLC,avg}_{j}$ define the average throughput in the eMBB slice and the average delay of the uRLLC slice for UEs in gNB $j$, and $D^{tar}$ refers to the target delay of the uRLLC slice. The allocation process ensures that each RB is assigned to a single UE, as stated in Equation $(\ref{eqn:2a})$, where $x_{j,u,r}$ and $x_{j,v,r}$ serve as binary variables denoting whether RB $r$ is designated for UE $u$ in eMBB slice and $v$ in uRLLC slice, respectively. Equation $(\ref{eqn:2b})$ indicates that the total number of RBs allocated to the gNB $j$ must not surpass the available resources. Finally, Equation $(\ref{eqn:2c})$ emphasizes that the assigned computation capacity to gNB $j$ should remain within the limits of the available resources.

Note that, in \cite{salehi2023adverserial,salehi2024jamming}, the transfer of RL-based knowledge from the expert to the learner agent was considered. However, in the current paper, we aim to examine the effect of transferring malicious DRL-based knowledge to the learner and subsequently evaluate the impact of the smart jamming attack and mitigation on the system.

    \vspace{-0.44cm}

\subsection{Knowledge Transfer by DTRL}
\label{subsection:DTRL}

Instead of starting from scratch, TL is utilized to share the knowledge of the expert with the learner. In \cite{zhou2022learning}, DTRL is proposed for resource allocation in the network slicing scenario which could be useful to represent complex policies, capturing patterns and representations across tasks. DTRL is also capable of tackling complex and high-dimensional state spaces \cite{yao2022tool}. The ability of DTRL to generalize knowledge from one task to another becomes imperative in this network-slicing scenario. 

The MDP definition of the DTRL for the expert and learner is shown in the Table. \ref{tab:merged}. 
In this definition $q^{eMBB}$ and $q^{uRLLC}$ represent the number of eMBB and uRLLC requests within the queue. 
The expert agent can assign RBs to eMBB and uRLLC UEs, shown by $r^{eMBB}$ and $r^{uRLLC}$. $r^{eMBB}$ and $r^{uRLLC}$ represent the radio RBs allocated to the eMBB and uRLLC slices respectively, available for both expert and learner agents. The learner agent can take into account joint resource allocation RBs and computation capability, then it has a different action set which is $(r^{eMBB},r^{uRLLC},c^{eMBB},c^{uRLLC})$ in which $c^{eMBB}$ and $c^{uRLLC}$ denote the computation capacity, exclusive to the learner agent. Similar to the states, the agents share a common reward function concerning meeting the requirements of slices.

\begin{table}[]
\caption{Transition tuples for the expert and learner users in the resource allocation algorithm}
\setstretch{1.2}
\resizebox{0.48\textwidth}{!}{
\begin{tabular}{|m{1.2cm}<{\centering}|m{8cm}<{\centering}|}
\cline{1-2}
\multicolumn{2}{|c|}{DTRL Algorithm for Network Slicing} \\ 
\cline{1-2}
State & $q^{eMBB},q^{uRLLC}$ \\ 
\cline{1-2}
Action (Expert) & {$r^{eMBB},r^{uRLLC}$} \\ 
\cline{1-2}
Action (Learner) & {$r^{eMBB},r^{uRLLC},c^{eMBB},c^{uRLLC}$} \\ \cline{1-2} 
Reward & $\omega^{eMBB}B^{eMBB,avg}+\omega^{uRLLC}(D^{tar}-D^{uRLLC,avg})$  \\ 
\cline{1-2}
\end{tabular}}
\label{tab:merged}
\vspace{-0.3cm}
\end{table}

Fig. \ref{fig:Sce} provides information about RBs allocation in the network slicing scenario by DTRL. In this system, the expert agent responsible for RB allocation evaluates the subsequent state and reward, as depicted in Equation (\ref{eqn:reward}), based on the feedback received from the system. The experience, $\{s_t, a_t, r_{t+1}, s_{t+1}\}$, is then stored in the experience replay memory to be used later for training the LSTM, where $s_t$, $a_t$, $r_{t+1}$, and $s_{t+1}$ are the state, action, reward, and next state, respectively. Afterward, the LSTM is used to predict the Q-values of all actions of the next state. 
Finally, the Q-values are fed to the $\epsilon-greedy$ algorithm for the next action selection. For the expert agent, the Q-table undergoes updates via:
\begin{multline}
      Q^{new}(s_t,a_t) = (1-\alpha)Q^{old}(s_t,a_t) + \\
      \alpha (r_{t+1} + \gamma \cdot \underset{a}{\max} Q(s_{t+1},a))),
            \vspace{-0.3cm}
\end{multline}
where $\alpha$ signifies the learning rate, $\gamma$ denotes the discount factor, $s_{t+1}$ embodies the next state after executing action $a_t$ in the state $s_t$. To incorporate the expert's knowledge into the learner's framework, the following equation, as detailed in \cite{zhou2022knowledge}, is employed:

\begin{multline}
      Q^{new}(s_{l,t},a_{l,t}) = Q^{T}(F(s_{l,t}),F'(a_{l,t})) + Q^{old}(s_{l,t},a_{l,t}) + \\
      \alpha (r + \underset{a}{\gamma\max} Q(s_{l,t+1},a)-Q^{old}(s_{l,t},a_{l,t}))),
\end{multline}
where $s_{l,t}$ and $a_{l,t}$ represent the current state and action taken by the learner agent at time $t$. The term $Q^{T}(F(s_{l,t}), F'(a_{l,t}))$ signifies the mapped Q-value, which guides the learner agent in using the expertise of the expert agent. Additionally, $F$ and $F'$ are mapping functions applied to the state and action, respectively, with $Q^{T}(F(s_{l,t}), F'(a_{l,t}))$ being influenced by the Q-value $Q^e(s_e, a_e)$, where $Q^e$, $s_e$, and $a_e$ correspond to the Q-value, states, and actions of the expert agent, respectively. The objective behind using $Q^{T}(F(s_{l,t}), F'(a_{l,t}))=Q^e(s, a)$ is to uncover a potential reward associated with the expert agent's actions that can be aligned with the learner's context. By determining the reward for an action in state $s_{l,t}$, the learner choose action $a_{l,t}$.

\section{Jamming Attacks on DTRL System}
\label{section:attacks}

The presence of jamming attacks in a wireless network poses challenges to the proper allocation of resources to users in network slicing. These attacks interrupt communication channels and disrupt them by inundating the channels with interference. When ML is used to control network slicing decisions, the momentary jamming effect might spread to longer time scales due to the inherent learning of ML-based schemes. In \cite{salehi2024jamming}, we examined the impact of CPJA on the TRL-based system, illustrating the degrading effect of the jamming attack on the RL-based resource allocation system. In this paper, we explore a DTRL-based scenario for radio resource allocation, then we evaluate the effect of the jamming attack on the system. Note that similar effects would be observed with other ML-based solutions however, we focus only on DTRL-based network slicing to keep the performance evaluations in a limited scope. This section examines several types of jamming attacks on the DTRL-based network slicing scenario.

\subsection{Constant Jamming Attack}
\label{subsection:CJA}

Wireless communication systems may be subject to constant jamming attacks (CJA) \cite{osanaiye2018statistical,salehi2024jamming} which aims to disrupt channels by continuous interference emitting. CJA has a significant impact on disrupting the flow of information between the gNB and UEs \cite{holtrup20215g}, leading to inefficient resource allocation, and impacting bandwidth distribution and network efficiency. Furthermore, if network slicing is implemented using ML, CJA can distort algorithm learning patterns, which results in suboptimal use of resources and degrades performance over the longer term.

For the CJA, the attacker selects the targeted $T^{th}$ $gNB$ for the attack and monitors the legitimate SINR of the gNB, denoted as $L$ calculated by the following equation:
\begin{equation}
\begin{aligned}
& L_{{j,u}} = \sum_{r \in N_u} \frac{p_{j,r}x_{j,u,r}q_{j,u,r}}{N_0+\sum_{j'\in J_{j'}}p_{j',r'}x_{j',u',r'}g_{j',u',r'}}, \\
\end{aligned}
\end{equation}
where $p_{j,r}$ stands for the transmission power allocated to RB $r$ within the gNB $j$, $x_{j,u,r}$ serves as a binary variable denoting whether RB $r$ is designated for UE $u$, and $q_{j,u,r}$ signifies the channel gain between gNB $j$ and UE $u$. $N_{0}$ characterizes the density of noise power, $J_{-j}$ offers insight into the aggregation of gNBs excluding the $j^{th}$ one, $u_{j'}$ embodies the congregation of UEs within gNB $j'$, and $N_{j'}$ symbolizes the entirety of RBs in gNB $j'$. The attacker then generates jamming signals with transmit power $P_{\psi_{CJA}}$ to interfere with communication between legitimate UEs and the associated gNB, which is calculated by the equation below:
\begin{equation}
    P_{\psi_{CJA}} = \frac{|\sum L_{j,u}|^2}{\lVert RBs\lVert}
\end{equation}
where a consistent pattern $P_{\psi_{CJA}}$ is established for generating deceptive signal transmissions during the attack, and $\lVert RBs\lVert$ shows the length of the RBs.
The subsequent step involves calculating the SINR of the targeted signal denoted as $\widehat{L}$ which is performed using the following equation:
\begin{equation}
\widehat{L}_{{j,u}} = \frac{p_{j,r}x_{j,u,r}q_{j,u,r}}{N_0+\sum_{j'\in J_{j'}}p_{j',r'}x_{j',u',r'}g_{j',u',r'}+P_{\psi_{CJA}}}, \\
\end{equation}

\begin{algorithm}[t!]
    \SetAlgoLined
    \textbf{Initialization:} Set network parameters and RB grid.\\
    \textbf{Expert Parameters:} $Q(s_t,\epsilon,a_t)$\\
    \textbf{Learner Parameters:} $Q_l(s_{l,t},\epsilon,a_{l,t})$\\
    \textbf{MDP Parameters:} ($S$,$A$,$R$,$\pi$,$\mu_0$)\\
    \For{\textbf{each TTI} $= 1$ to $T$}{
        \For{\textbf{each RB} $= 1$ to $RBG$}{
            Target $T^{th}$ gNB for CJA\\
            Chooses a subcarrier of the $T^{th}$ gNB \\ 
            Calculate $L_{j,u}$ of the $T^{th}$ gNB \\ 
            Generate perturbing signal $P_{\psi_{CJA}}$ \\
            Transmit $P_{\psi_{CJA}}$ in TTIs 
            Receiving $\widehat{L}_{j,u,\psi_{d}}$ \\
            Effecting RBs allocation of  $Q^{new}(s_t,a_t)$\\
            \[ Q_{CJA}^{new}(s_t,a_t) \]\\
            Update expert policy to: \\
            \[ Q_{CJA}^{new}(s_t,a_t) \]\\
            Update learner's knowledge to: 
            \[ Q_{l,CJA}^{new}(s_{l,t},a_{l,t}) \]
        }
    }
    \caption{Procedure for Continuous Jamming Attack (CJA) on DTRL Knowledge Base.
}
    \vspace{-0.1cm}
\end{algorithm}
\textit{Algorithm 1} explains the implementation of CJA on the DTRL-based network slicing, where TTI represents the transmission time interval. In the CJA, $P_{\psi_{CJA}}$ disrupts the DTRL model's learning patterns, distorting system capabilities since $P_{\psi_{CJA}}$ compromises the expert agent's decision-making for RB allocation. This results in suboptimal choices that reduce the reliability and efficiency of the system. Upon poisoning the expert agent's knowledge, the jamming influences the RB allocation of the learner agent through knowledge transfer. The resulting updated values, $Q_{{CJA}}^{new}(s_t,a_t)$, are then transferred to the learner agent as $Q_{l,{CJA}}^{new}(s_{l,t},a_{l,t})$, reflecting the impact of the jamming on the learning process.

While CJA impacts the system's performance and poses a vulnerability to DTRL, monitoring the system enables the detection of this attack. Consequently, attackers attempt to alter their transmission power to avoid detection and amplify their influence randomly. Subsection \ref{subsection:RJA} and \ref{subsection:DRL-JA} detail the procedure of the random and intelligent attackers employing random and adaptive jamming attacks on the system.

\subsection{Random Jamming Attack}
\label{subsection:RJA}
One of the other common types of attacks is the random jamming attack (RJA). It should be noted that the transmission power of the attacker during the RJA ($P_{\psi_{RJA}}$) is the same as $P_{\psi_{CJA}}$ which depends on the $T^{th}$ gNB. However, unlike the CJA, in the RJA, the jammer randomly transmits jamming signals, affecting a random number of RBs.

\subsection{Intelligent DRL-based Jamming Attack}
\label{subsection:DRL-JA}
\vspace{-0.135cm}
As system security improves and attacks are detected more efficiently, attackers employ intelligent and ML-based strategies to evade detection. These attacks dynamically adjust interference patterns in response to network conditions, reducing the effectiveness of conventional defense mechanisms. Our adaptive jamming attack by DRL algorithm (DRL-JA) monitors the $T^{th}$ gNB and generates interference accordingly.

\begin{figure*}[t!]
  \centering
  \includegraphics[width=1\linewidth]{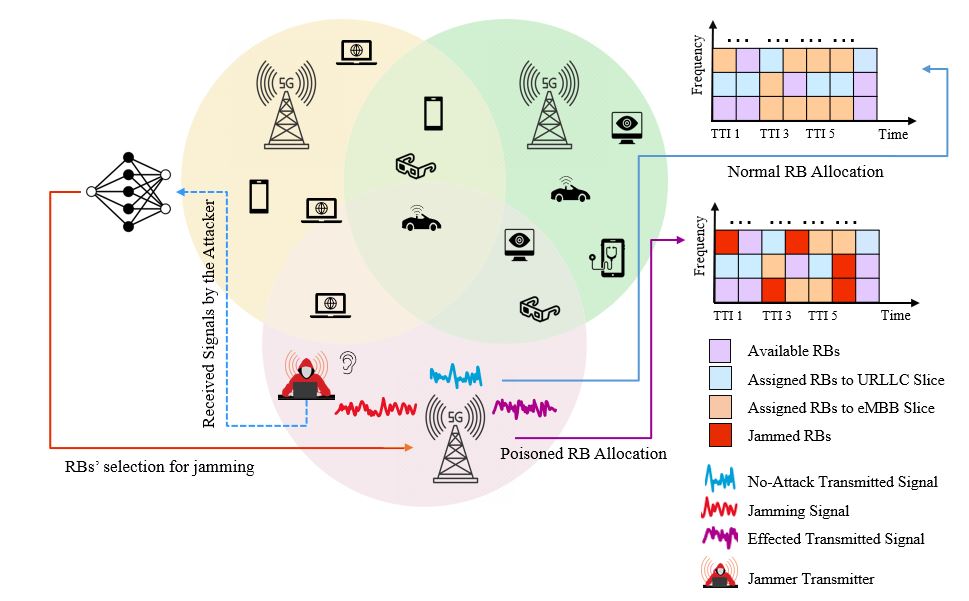}
  \caption{Jamming attacks on the network slicing scenario 
 }
    \label{fig:att}
    \vspace{-0.25cm}
\end{figure*}

As depicted in Fig. \ref{fig:att}, during the DRL-JA, the intelligent attacker overhearing the subcarriers of $T^{th}$ gNB to gather information about the RBs allocation, to intentionally interfere, manipulate, and perturb resource allocations. To accomplish this, prior research has examined the case of an attacker that has access to a physical downlink control channel (PDCCH) to jam systems \cite{arjoune2020smart,debruhl2011digital}. However, we aim to examine the situation where the jammer does not have access to the PDCCH. Then in the DRL-JA, the attacker employs spectrum sensing to evaluate the channel allocation to UEs. In this evaluation, the fundamental unit is a sensing slot ($T_{sl}$), which lasts for $T_{sl} = 9us$ \cite{3gppstandard}. As long as the received power is below $E_{\theta}$ for at least $4us$ during the sensing, the attacker considers $T_{sl}$ to be idle, otherwise, the slot is considered as the allocated. During the DRL-JA, the attacker targeted the allocated slots and transmitted interference signals on those slots. It must be noted that as the DRL-JA operates with limited energy for channel interference, it actively monitors the transmission power of the $T^{th}$ gNB ($P_{T^{th}}$) to detect assigned RBs. 

To execute DRL-JA, the attacker monitors the system, modeled by the MDP. The MDP of DRL-JA is denoted as $M_{DRL-JA} = (s_{DRL-JA}, a_{DRL-JA}, r_{DRL-JA})$. The state ($s_{DRL-JA}$), action ($a_{DRL-JA}$), and reward for the DRL-JA ($r_{DRL-JA}$) will be defined in the following.
\begin{itemize}
\item \textbf{State}: For $s_{DRL-JA}$, the attacker utilizes $P_{T^{th}}$ in the estimation of RBs allocation, as shown in the following equation:
\begin{equation}
    P_{T^{th}} =
    \begin{cases}
       (1 - \vartheta)P_{\psi_{DRL-JA}}, &  P_{T^{th}} > E_{\theta} \\
      \vartheta P_{\psi_{DRL-JA}} , &   P_{T^{th}} \le E_{\theta}  
    \end{cases} 
\end{equation}
where $\vartheta$ is a binary variable that acts as a switch that determines which term contributes to the value of $P_{T^{th}}$ based on the condition specified in the state definition. In calculating the $E_{\theta}$ for the transmission signal, the following equation is employed:
\begin{equation} 
    E_{\theta} = 
      \min
      \begin{cases}
        T_{\max}+10\text{dB}, \\
        X_r,
      \end{cases}
\end{equation}
where $X_r$ is the maximum energy detection threshold specified according to regulatory requirements in $\text{dB}$, and $T_{\max}$ is determined by the equation below:
\begin{multline}
   T_{\max}(\text{dBm})= \\
   10 \cdot \log_{10} (3.16228 \cdot 10^{-8}\left(\frac{mW \, }{ \, \text{MHz}}\right)\cdot b_q (\text{MHz})), \, 
\end{multline}
where $b_q$ is the bandwidth of RB $q$.
\item \textbf{Action}: The jammer's action set, denoted as $a_{DRL-JA}$, comprises a finite set of actions employed to simulate interference on the allocated RBs within a subcarrier. The actions set is defined by the following equation:
\begin{equation}
a_{DRL-JA} = \sum_{r \in N_u} \widetilde{P}_r,
\end{equation}
where $\widetilde{P}_r$ signifies the interference power transmitted by the jammer on the number of RBs, and $P_{\psi_{DRL-JA}}=\widetilde{P}_r$.
\item \textbf{Reward}: The reward function is a combination of two key components, as depicted in the equation below:
\begin{equation}
    r_{DRL-JA}=\omega_1 \widetilde{L}+\omega_2 \widetilde{E},
\end{equation}
where \(\omega_1\) serves as a weight to prioritize interference control, and \(\widetilde{L}\) reflects the \(SINR\) of the system due to DRL-JA, which is calculated by the equation below:
\begin{equation}
 \widetilde{L}_{{j,u}} = \frac{p_{j,r}x_{j,u,r}q_{j,u,r}}{N_0+\sum_{j'\in J_{j'}}p_{j',r'}x_{j',u',r'}g_{j',u',r'}+P_{\psi_{DRL-JA}}}, \\
\end{equation}
The second term, \(\omega_2 \widetilde{E}\), considers energy-related aspects, where \(\omega_2\) is a weight and \(\widetilde{E}\) represents energy consumption metrics. The energy consumption during the DRL-JA is simplified by:
\begin{equation}
    \widetilde{E} = P_{\psi_{DRL-JA}}\cdot T_{DRL-JA},
\end{equation}
where $T_{DRL-JA}$ is the duration of the DRL-JA transmission. This composite reward metric aims to strike a balance between optimizing interference mitigation and promoting energy efficiency, thus guiding the attacker toward more effective and resource-aware jamming strategies.
\end{itemize}

Employing a DRL algorithm to observe the dynamics of $T^{th}$ gNB, the attacker utilizes the policy function ($\pi$), which maps states to actions and shapes the attacker's behavior within this evolving environment, can be defined as:
\begin{multline}
    Q_{DRL-JA}^{\pi} (s_{DRL-JA},a_{DRL-JA}) = E_{\pi} \\
    \Biggl [\sum_{t=0}^{TTI} \gamma_{DRL-JA}^{t} r_{{t}_{DRL-JA}}|s_{DRL-JA}^{0} = s_{DRL-JA}^{ }, \\
    a_{DRL-JA}^{0} = a_{DRL-JA}^{ } \Biggl],
    \vspace{-0.3cm}
\end{multline}
where $\gamma_{DRL-JA}$ is the discount factor. The attacker by the MDP aims to discover an optimal policy, denoted by $\pi_{DRL-JA}^*$, which corresponds to a value function $Q_{DRL-JA}^{*} (s_{DRL-JA}, a_{DRL-JA})$.

The DRL-JA utilizes spectrum sensing to detect transmission patterns then builds knowledge based on observed RB allocations over time, learning which RBs are likely to be used. After monitoring the environment and acting, the DRL-JA refines its policy using RL which relies on its perception of how its actions influence the system's behavior. It should be noted that the DRL-JA has limited power resources. It must balance energy consumption and interference efficacy. This constraint forces it to make strategic decisions about when and where to jam. The DRL-JA's physical location is considered static.

\section{Intelligent DRL-based Attack Mitigation}
\label{section:Mitigation}

In this section, we introduce an intelligent mitigation technique for the DRL-JA and CJA. 

\subsection{DRL-based Jamming Attack Mitigation}
\label{subsection:DRL_Mitigation}

\begin{figure*}[t!]
  \centering
  \includegraphics[width=1\linewidth]{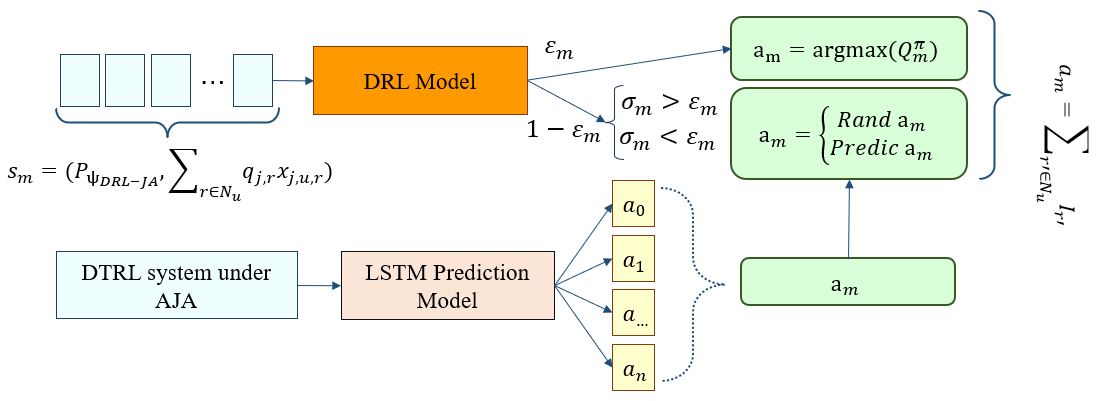}
  \caption{DRL-based Model of Intelligent-DRL-based Mitigation 
  }
    \label{fig:mit}
    \vspace{-0.3cm}
\end{figure*}

In the DRL-based attack mitigation, as depicted in Fig. \ref{fig:mit} the security administrator mitigates the activity of the attacker by monitoring. The security administrator by accessing the $a_{DRL-JA}$ and $\Sigma_{i=1}^{q}\Sigma_{s=1}^{sub} RB_{q,s}$, which illustrate the transmission signal of the attacker on the selected RBs and the expert agents RB allocation, try to mislead the attacker by the generated signals on unallocated RBs. The mitigation method by deep Q-learning algorithm uses experience replay to break the correlation between consecutive experiences, resulting in more stable and effective training. It is also able to learn from a wider range of experiences, thus improving generalization and learning capabilities. Then the administrator by using the DRL model generates signals on the vacant RBs and the attacker targets them for its attack. 

The DRL transition tuples for the DRL-based mitigation are characterized by states, actions, and the reward component, as given below: 
\begin{itemize}
\item \textbf{State}: The mitigation method's state ($s_m$) encompasses details regarding the jammed RBs on their corresponding subcarriers by $P_{\psi_{DRL-JA}}$, and the allocated RBs on the subcarriers in the expert agent which is depicted by $\sum_{r \in N_u} q_{j},x_{j,u,r} $. By analyzing both the RB allocation policy data and the attacker's jamming signals, the security administrator gains a more comprehensive understanding of the system. This enhanced perspective allows for smarter decisions to mislead the attacker and better defend the network. 

\item \textbf{Action}: The action set ($a_m$) is determined by the summation of interference powers on unallocated RBs ($r'$). Mathematically, this is represented by the equation below:
\begin{equation}
  a_m = \Sigma_{r' \in N_u} \mathbb{I}_{r'}, 
\end{equation}
where $\mathbb{I}$ represents the interference power of the considered antenna when transmitting signals on $r'$. For the action selection in the mitigation method, the Q-values are fed to the $\epsilon-greedy$ algorithm.

\item \textbf{Reward}: In the mitigation method, the security administrator tries to increase the probability of miss detection of assigned RB in the attacker. Miss detection occurs when the attacker is unable to identify allocated RBs. The reward ($r_m$) is computed based on the average SINR of the system during the mitigation method. This reward metric serves as an indicator of the quality of the communication channel in the presence of the attack. 
\begin{equation}
    r_{m} = \overline {L_{{j,u}}}
\end{equation}
\end{itemize}

The security administrator by the MDPs aims to discover an optimal policy, denoted by $\pi_{m}^*$, which corresponds to an expected return $V_{m}^*(s_{m})$ and a value function $ Q_{m}^{*} (s_{m}, a_{m})$. In the context of DRL, the Q-function follows the Bellman optimal equation. 
The security administrator uses deep Q-learning to estimate Q-values based on rewards and the Q-value function, which follows the equation below:

\begin{multline}
      Q^{'}_{m}(s_{m},a_m) = Q_m(s_{m},a_m) + \\
      \chi (r_m + \gamma \underset{a'_m}{\max} Q(s'_{m},a'_m)-Q^{old}(s_{m},a_m))),
              \vspace{-0.3cm}
\end{multline}
where $\chi$ is the learning rate. 

The mitigation system assumes knowledge of the attacker’s jamming behavior by monitoring DRL-JA’s actions on allocated RBs. It requires feedback on SINR and system performance after the mitigation strategy is applied. For optimal performance, real-time data on jammed RBs, system RB allocation, and interference signals is necessary. The output involves sending misleading signals on unallocated RBs and measuring SINR to evaluate improvements. The system uses power resources for interference on vacant RBs, without changing channel usage, and assumes sufficient energy availability for this purpose.

In the mitigation method, the additional antenna requires energy. The power consumption of the antenna, $P_{DRL_{mit}}$, for generating interference signals can be expressed as:
\begin{equation}
    P_{DRL_{mit}} = \sum_{i=1}^{N} p_i \cdot \mathbb{I}_{\{i \in \mathfrak{B}\}}
\end{equation}
where $p_i$ is the power required to generate an interference signal on the $i_{th}$ RB, $N$ is the total number of RBs, $\mathfrak{B}$ is the set of RBs targeted for transmitting signals on, and $\mathbb{I}_{i \in \mathfrak{B}}$ is an indicator function that equals 1 if the $i_{th}$ RB is in the set $\mathfrak{B}$, and 0 otherwise. $p_i$ would be the power of the signal $\xi[n]$, which is generated by the antenna for misleading the jammer. Then the average power can be calculated as:
\begin{equation}
  p_i = \frac{1}{N} \sum_{n=0}^{N-1}|\xi[n]|^2
\end{equation}
where $N$ is the number of signals generated by the antenna for misleading the DRL-JA.

While this method requires extra resources, some methods, such as \cite{shi2021attack,erpek2018deep,sharma2022mitigating}, do not use additional antennas or channels, resulting in lower $p_i$ values but greater susceptibility to jamming attacks. These approaches make the system more vulnerable to intelligent jamming, making them suitable for energy-limited systems but less effective in high-threat environments. By using extra antennas or channels to transmit misleading signals, the proposed method increases channel usage and energy consumption but enhances resilience against attacks. Its advantage lies in a more adaptive, deceptive strategy that offsets additional resource consumption by ensuring continued system performance in adversarial scenarios. The main benefit is its ability to mislead the attacker, improving overall system performance despite increased energy use. In this paper, instead of focusing on power balancing, the comparison is enhanced by considering system performance metrics such as throughput improvements and latency. 

\subsection{Jamming Attacks Mitigation by Learning Suspending}

In order to mitigate jamming attacks, modifications have been made to the baseline model presented in \cite{shi2021attack} to prevent the system from learning if the system criteria fall below a certain threshold. The threshold, dependent on $P_{T^{th}}$ and interference $L{{j,u}}$, which are calculated under no-attack situations. To obtain $(\overline{P_{T^{th}}},\overline{L{{j,u}}})$, the system is monitored in 10 runs, and the averages of 5 runs are calculated to evaluate these parameters in a no-attack scenario. This information is stored in the knowledge repository ($T_{KB}$), and during the DTRL approach if an attack is detected, the learning process is suspended, and the previous system action ($a'_{t-1}$) is selected. When the current system state ($s'_t$) does not fall below the threshold, the optimal action is chosen. \textit{Algorithm 2} provides details about the baseline attack mitigation. It's worth noting that the system could be affected by either $P{\psi_{CJA}}$ or $P_{\psi_{DRL-JA}}$, denoted by $(\widehat{L} \vee \widetilde{L})$.

\begin{algorithm}[t!]
    \caption{Suspended Learning Attack Mitigation}
    \SetAlgoLined
    \DontPrintSemicolon
    \textbf{Input:} Knowledge repository based on the secure system, $T_{KB} = (\overline{P_{T^{th}}},\overline{L_{{j,u}}})$.\\
    Initialization\;
    \For {t $=$ 1 to TTI} {
        Monitor the system state: $s'_{t} = (P_{T^{th}}, \widehat{L} \vee \widetilde{L})$\;
        \If {$T_{KB}[1]<s'_{t}[1]$ and $T_{KB}[2]>s'_{t}[2]$}
        {
            Attack detection\;
            Suspend system updating and revert to the knowledge $(s'_{t-1},a'_{t-1})$\;
        }
        \Else {
            No attack found\;
            The system is updated by $(s'_{t},a'_{t})$\;
        }
    }
    \label{alg:PoEG}
    \vspace{-0.1cm}
\end{algorithm}

\section{PERFORMANCE EVALUATION}
\label{section:result}

\subsection{Parameter Settings}

In this scenario, three gNBs have been deployed, each providing a coverage radius of 125 meters and boasting a bandwidth of 20 MHz, with 13 RB groups (RBGs) at their disposal. In the considered system, there are 12 subcarriers, and the number of RBGs is equal to 8. Within each gNB, network slices have been allocated to cater to uRLLC and eMBB services, accommodating 10 to 40 UEs. For our analytical investigation, we consider uRLLC packets 50 bytes in size, while eMBB packets are set at 100 bytes \cite{zhou2022knowledge}. To capture traffic dynamics intricacies, we employ a data generation model based on a Poisson distribution. Our computational infrastructure in this context operates at a clock speed of 1 GHz, allowing for 200 CPU cycles per bit for processing tasks.

\begin{table}[h!]
\caption{Simulation Settings}
\centering
\small
\resizebox{0.48\textwidth}{!}{
\begin{tabular}{|m{2.6cm}<{\centering}|>{\raggedright\arraybackslash}p{5.5cm}|}
\hline
Parameters & Requirements constraint \\ \hline
Network & 3GPP Urban Macro (UMa) network \\
Environment                & 5 gNodeBs and 500 meter inter-site distance \\ \hline
 & 15 kHz subcarrier spacing \\
                           & 12 subcarriers per RB \\
                           & Number of RBG K = 13 \\
         & TTI size of 2 OFDM symbols (0.1429 ms) \\
PHY configuration         & Max. transmission power of 40 dBm [13] \\
                           & Tx/Rx antenna gain of 15 dB \\ 
                           & Bandwidth = 20 MHz \\
                           & Carrier frequency of 30 GHz \\ \hline
& Asynchronous HARQ \\
              & Round trip delay is 4 TTIs \\
HARQ          & 6 HARQ processes \\
              & Max. 1 HARQ re-transmission \\ \hline
 & $128.1 + 37.6 log_{10}{(D[Km])}$ \\
Propagation      & Log-Normal Shadowing (8 dB) \\
                & Noise Figure of 5 dB \\
                & Penetration loss of 5 dB \\ \hline
 & Stationary and uniformly distributed \\
User distribution        & [10 : 10 : 30] uRLLC \\
                           & 25 eMBB \\ \hline
 & uRLLC: 20\% CBR and 80\% Poisson \\
Traffic model       & eMBB: Poisson \\
                       & Payload size: 32 Byte \\ \hline
uRLLC Load/Cell & [1 : 1 : 4] Mbps \\ 
eMBB Load/Cell & [1 : 1 : 4] Mbps \\ \hline
 &     Size of input layer = 1 \\
                &     Number of hidden units = [10: 10: 40] \\
                &     Size of output layer = 13 \\
LSTM Parameters  &     Size of mini-batch = 20 \\
                &     Size of replay memory = 60 \\
                &     Training Interval = 60 \\
                &     Copy Interval = 120 \\ \hline
  & Learning rate $\alpha$ = 0.5 \\
Q-learning      &       Discount factor $\gamma$ = 0.9 \\
                &       Exploration probability $\epsilon$ = 0.1 \\ \hline
Optimization    & $w^{eMBB}$ = 1 \\
Parameters      &       $w^{uRLLC}$ = 1 \\ \hline                
\end{tabular}}
    \label{tab:parameters}
\end{table}

\subsection{Accelerated Convergence through DTRL}

\begin{figure}[t]
\centering
\begin{subfigure}{0.48\textwidth}
    \centering
    \includegraphics[width=0.87\textwidth]{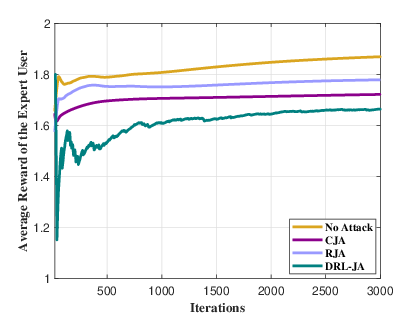}
    \caption{Reward convergence for Expert agent}
    \label{fig:ConvEx}
\end{subfigure}
\hfill
\begin{subfigure}{0.48\textwidth}
    \centering
    \includegraphics[width=0.87\textwidth]{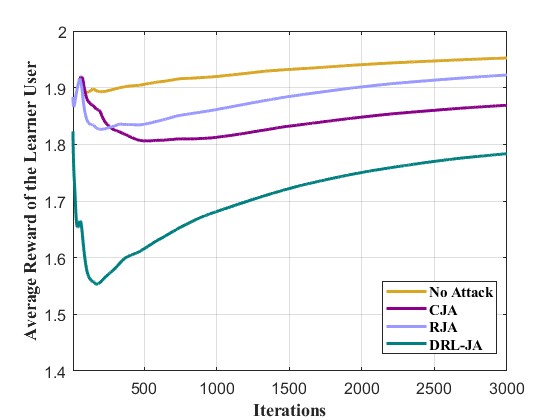}
    \caption{Reward convergence for Learner agent}
    \label{fig:ConvLe}
\end{subfigure}
\caption{Average reward convergence under various attacks}
\end{figure}

In the considered scenario, the DTRL algorithm proves its effectiveness by utilizing existing knowledge to address data scarcity and the need for accurate, adaptable models. According to Fig. \ref{fig:ConvEx} and Fig. \ref{fig:ConvLe}, both expert and learner rewards fluctuate during initial iterations, but the expert reward converges after approximately 1,000 iterations without attack, while the learner's reward converges after approximately 500 iterations. A significant advantage of integrating DTRL into network slicing is its ability to speed up the convergence of learner agents within the network. Furthermore, in Fig. \ref{fig:ConvEx} and Fig. \ref{fig:ConvLe}, the system reward under attacks is depicted, showing that the RJA has less impact on the system compared to the CJA. The consistent interference of CJA disrupts uRLLC service requests significantly because these services depend on a highly predictable and stable communication channel. CJA also affects throughput consistently, leading to a steady degradation in service quality. It should be noted that as eMBB can tolerate more latency and slight reductions in reliability, the impact, while noticeable, is not as considerable as in uRLLC. While RJA is disruptive, its stochastic interference is less affected, potentially allowing for some transmission during less jammed intervals. However, RJA's unpredictability poses a challenge to maintaining the required throughput and latency for eMBB and uRLLC slices. As CJA is more vulnerable against the DTRL system, we consider the effect of the CJA on the system, and evaluate the system performance under this attack. 

Taking into account the impact of the attacks, the likelihood of convergence within the first 1000 iterations of the system model decreases, and the attacks reduce both agents' system rewards. Furthermore, the DRL-JA has a more considerable impact on the system's perturbations as the adversary perturbed the assigned RBs to UEs. Despite DTRL's efficacy in propagating knowledge across diverse scenarios, this method transfers malicious knowledge. Consequently, it highlights the importance of considering attack mitigation methods in TL-based methods. 

\subsection{CJA and DRL-JA Impact on DTRL System}

\begin{figure}[t]
\centering
\begin{subfigure}{0.48\textwidth}
    \centering
    \includegraphics[width=0.87\textwidth]{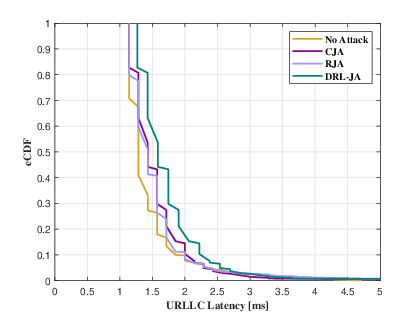}
    \caption{eCDF for Expert agent under various attacks}
    \label{fig:ExeCDF}
\end{subfigure}
\hfill
\begin{subfigure}{0.48\textwidth}
    \centering
    \includegraphics[width=0.87\textwidth]{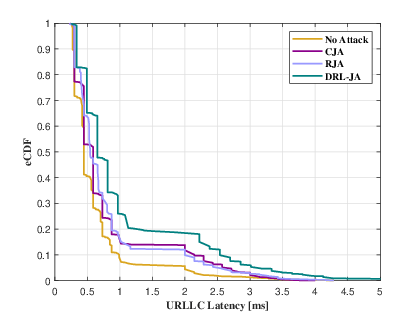}
    \caption{eCDF for Learner agent under various attacks}
    \label{fig:LeeCDF}
\end{subfigure}
\caption{System eCDF under the attacks}
\end{figure}

Fig. \ref{fig:ExeCDF} and Fig. \ref{fig:LeeCDF} illustrate the empirical cumulative distribution function (eCDF) representing the latency for uRLLC in four distinct scenarios no attack, CJA, RJA, and DRL-JA. As depicted in Fig. \ref{fig:ExeCDF} and Fig. \ref{fig:LeeCDF}, the RJA has less impact on the system than the CJA. In CJA, a constant amount of interference affects the system, reducing its performance in meeting uRLLC and eMBB requirements. It is evident from this graphical representation that the system behaves differently in the presence of jamming and absence, thus providing insight into the degradation sustained by the system. Notably, the findings for no attack cases highlight that the expert case exhibits the highest latency distribution. This difference is attributed to the absence of MEC server deployment within the expert and the processing computation tasks in the central cloud infrastructure. Consequently, this centralized approach leads to elevated latency levels.

It is clear that DRL-JA has a more vulnerable effect on the system and leads to more interference in the system, whereas in contrast, the CJA effect does not have such a great impact on the system. This depiction helps quantify the DRL-JA's influence and the necessity to formulate mitigation to boost the DTRL system's robustness and resilience.

\begin{figure*}
\centering
\includegraphics[width=0.95\textwidth]{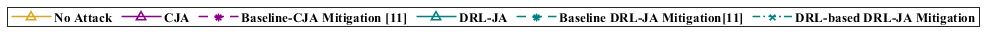}
\label{fig:Main_Figure}
\vspace{0.001em} 

\begin{subfigure}{0.32\textwidth}
    \includegraphics[width=\textwidth]{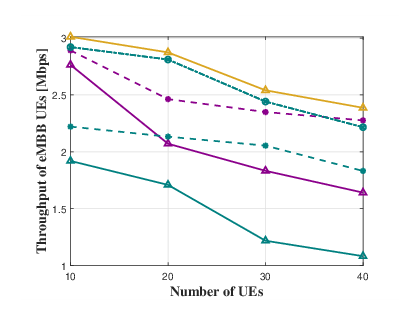}
    \caption{Throughput of eMBB slice under different UEs}
    \label{fig:Ex_Th_UE}
\end{subfigure}
\hfill
\begin{subfigure}{0.32\textwidth}
    \includegraphics[width=\textwidth]{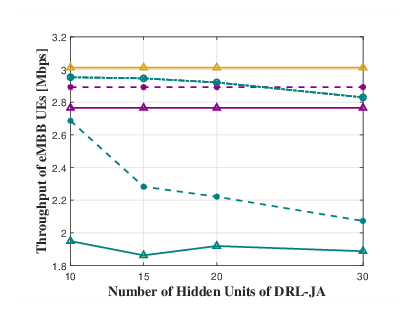}
    \caption{Throughput under different hidden units of DRL-JA}
    \label{fig:Ex_Th_HU}
\end{subfigure}
\hfill
\begin{subfigure}{0.32\textwidth}
    \includegraphics[width=\textwidth]{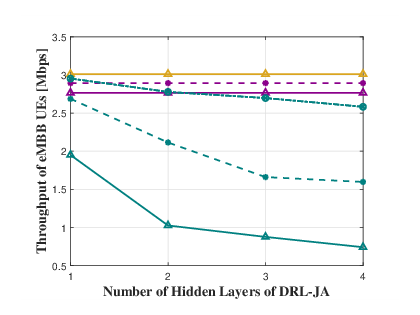}
    \caption{Throughput under different hidden layers of DRL-JA}
    \label{fig:Ex_Th_HL}
\end{subfigure}

\caption{Throughput of the Expert agent under the attacks and mitigations.}
\label{Expert_Th}
\vspace{-15pt}
\end{figure*}

\begin{figure*}
\centering
\begin{subfigure}{0.32\textwidth}
    \includegraphics[width=\textwidth]{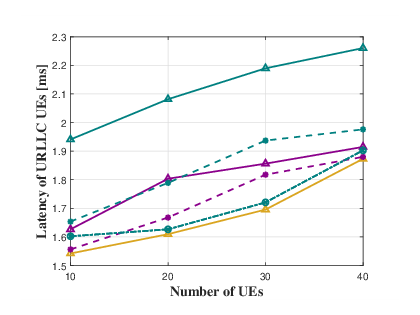}
    \caption{Latency of uRLLC slice under different UEs requirments}
    \label{fig:Ex_Le_UE}
\end{subfigure}
\hfill
\begin{subfigure}{0.32\textwidth}
    \includegraphics[width=\textwidth]{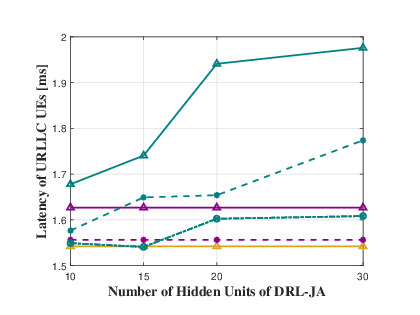}
    \caption{Latency under different hidden units of DRL-JA}
    \label{fig:Ex_Le_HU}
\end{subfigure}
\hfill
\begin{subfigure}{0.32\textwidth}
    \includegraphics[width=\textwidth]{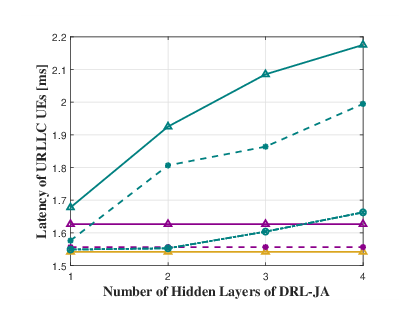}
    \caption{Latency under different hidden layers of DRL-JA }
    \label{fig:Ex_Le_HL}
\end{subfigure}
\caption{Latency of the Expert agent under the attacks and mitigations.}
\label{Expert_Le}
\end{figure*}

\begin{figure*}
\centering
\includegraphics[width=0.95\textwidth]{Fig/Legend.JPG}
\label{fig:legend}
\vspace{0.001em} 
\begin{subfigure}{0.32\textwidth}
    \includegraphics[width=\textwidth]{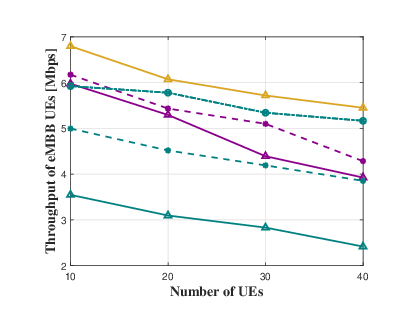}
    \caption{Throughput of eMBB slice under different UEs}
    \label{fig:Le_Th_UE}
\end{subfigure}
\hfill
\begin{subfigure}{0.32\textwidth}
    \includegraphics[width=\textwidth]{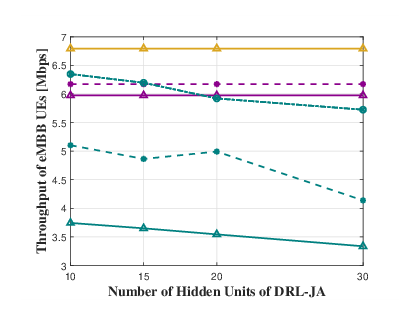}
    \caption{Throughput under different hidden units of DRL-JA}
    \label{fig:Le_Th_HU}
\end{subfigure}
\hfill
\begin{subfigure}{0.32\textwidth}
    \includegraphics[width=\textwidth]{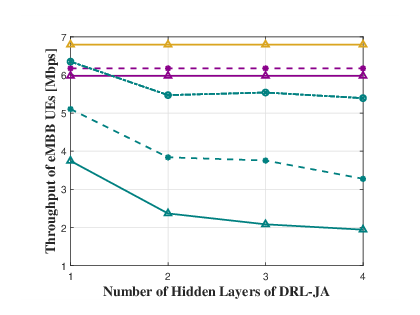}
    \caption{Throughput under different hidden layers of DRL-JA}
    \label{fig:Le_Th_HL}
\end{subfigure}
\caption{Throughput of the Learner agent under the attacks and mitigations.}
\label{Learner_Th}
\vspace{-15pt}
\end{figure*}

\begin{figure*}
\centering

\begin{subfigure}{0.32\textwidth}
    \includegraphics[width=\textwidth]{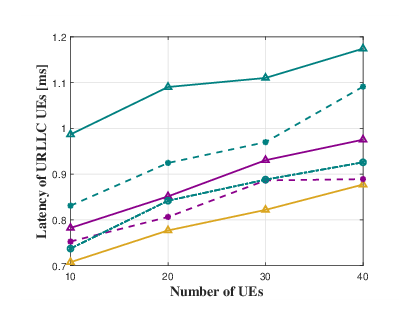}
    \caption{Latency of uRLLC slice under different UEs requirements}
    \label{fig:Le_Le_UE}
\end{subfigure}
\hfill
\begin{subfigure}{0.32\textwidth}
    \includegraphics[width=\textwidth]{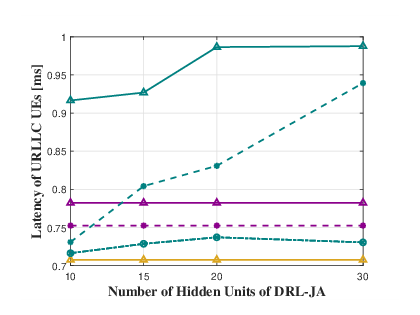}
    \caption{Latency under different hidden units of DRL-JA}
    \label{fig:Le_Le_HU}
\end{subfigure}
\hfill
\begin{subfigure}{0.32\textwidth}
    \includegraphics[width=\textwidth]{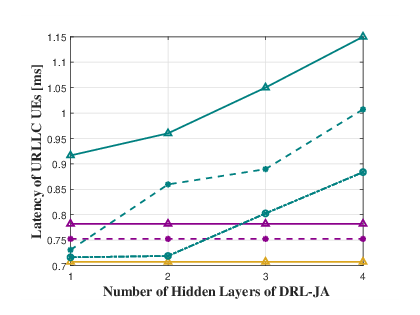}
    \caption{Latency under different hidden layers of DRL-JA}
    \label{fig:Le_Le_HL}
\end{subfigure}

\caption{Latency of the Learner agent under the attacks and mitigations.}
\label{Learner_Le}
\end{figure*}

Fig. \ref{fig:Ex_Th_UE}, Fig. \ref{fig:Ex_Le_UE}, Fig. \ref{fig:Le_Th_UE}, and Fig. \ref{fig:Le_Le_UE} illustrate the system throughput and latency for the expert and the learner in three distinct scenarios: no attack, CJA, and DRL-JA. System key performance indicators (KPIs) are assessed across various numbers of UEs, ranging from 10 to 40.
As shown in Fig. \ref{fig:Ex_Th_UE} and Fig. \ref{fig:Le_Th_UE} due to the difficulty of the algorithm in distinguishing sporadic interference bursts, the DRL-JA is more likely to disrupt system throughput than the CJA, resulting in a greater reduction in performance. Increasing the traffic load leads to reducing the throughput of the system in scenarios with no attack and under attacks. Regarding the effect of the attacks on the system, the jammer by using DRL-JA perturbs the system throughput more severely than CJA. Following Fig. \ref{fig:Ex_Le_UE} and Fig. \ref{fig:Le_Le_UE}, the DRL-JA also result in increased latency in systems that utilize DRL algorithms. Due to the variable interference patterns of the DRL-JA, the DTRL algorithm cannot adapt quickly, leading to fluctuations in communication delays and overall latency of both agents. It is important to note that one of the primary causes of increased latency is signal disruption caused by jamming, which hinders the timely transmission of critical data between the gNB and UEs. Furthermore, CJA and DRL-JA can cause packet loss, corruption, or delay as a result of interference with communication channels. Therefore, the DTRL system is experiencing delays in receiving essential updates, model parameters, or feedback, which are pivotal to its learning and decision-making processes. There is also a need for additional error detection and correction mechanisms in CJA and DRL-JA, which contributes to heightened latency. To deal with corrupted or lost data caused by jamming, the DTRL system implements retransmissions, all of which consume valuable time and resources.

Fig. \ref{fig:Ex_Th_HU} and Fig. \ref{fig:Ex_Le_HU} for the expert agent and Fig. \ref{fig:Le_Th_HU}, Fig. \ref{fig:Le_Le_HU} provide more information about the effect of utilizing various hidden units by the DRL-JA on the system performance. As depicted in these figures, by increasing the number of hidden units, the DRL-JA's effect increases on the expert agent, and subsequently, it increases on the learner agent as well. With more hidden units, the jammer can predict the allocation of RBs more accurately and affect the system more significantly. Fig. \ref{fig:Ex_Th_HL} and Fig. \ref{fig:Ex_Le_HL} also illustrate the effect of using more hidden layers on the DRL-JA's impact on the expert agent, and Fig. \ref{fig:Le_Th_HL} and Fig. \ref{fig:Le_Le_HL} show this effect on the learner agent. As depicted in these figures, by increasing the number of hidden layers, the jammer's effect on the system increases for both KPIs. Increasing the number of hidden units and layers enhances the neural network's capacity to learn complex patterns and make more accurate predictions, allowing the DRL-JA to more effectively disrupt RB's allocation. This improved capability leads to a greater impact on both the expert and learner agents' performance.

Jamming attacks adversely affect the DTRL system's throughput because they disrupt the seamless flow of packets. This decrease in throughput can be attributed primarily to the interference caused by the $P_{\psi_{CJA}}$ and $P_{\psi_{DRL-JA}}$, which disrupt the transmission of data packets between the $T^{th}$ gNB and the UEs. Jamming-induced packet losses necessitate retransmissions, consuming valuable resources and time. Additionally, the continuous presence of $\psi_{CJA}$ prompts agents to adopt conservative transmission strategies to enhance robustness to interference. These adaptations, while mitigating the impact of jamming in CJA, inevitably lead to a decrease in achievable data rates, directly impacting throughput. It is also unfortunate that the discontinuous presence of $\psi_{DRL-JA}$ causes expert agents to be unable to adopt that information and that unstable data are transmitted to learners. The detrimental influence of DRL-JA on the DTRL system demonstrates itself as a visible decrement in network throughput.

In the context of the DTRL system, the impact of different CJA and DRL-JA can have varying effects on system latency. Firstly, DRL-JA is characterized by its adaptability and ability to target critical network resources selectively. When DRL-JA is launched on the DTRL system, it increases latency. The reason for this is that the intelligent jammer is adept at identifying and disrupting key RBs. Targeting the most critical components of the DTRL system, they disrupt the normal flow of information, leading to delays in data transmission, processing, and decision-making. These disruptions result in heightened latency, which negatively impacts system performance and responsiveness. On the other hand, CJA operates differently, emitting continuous interference signals across a wide spectrum, affecting RBs randomly. While CJA may not be as sophisticated as DRL-JA, it has a substantial impact on system latency. Continuous interference generated by constant jammers introduces noise and contention for network resources. This noise leads to packet collisions, retransmissions, and increased queueing delays, all of which contribute to higher latency in the DTRL system. Overall, both CJA and DRL-JA have effects on various aspects of the DTRL system and can be summarized as follows:
 \begin{itemize}
    \item \textbf{False Learning Signals}: The CJA and DRL-JA introduce interference, causing the expert agent in the network slicing system to make incorrect observations. This interference affecting the accuracy of $Q^{new}(s_t,a_t)$ updates, which are affecting $Q^{new}(s_{l,t},a_{l,t})$ due to the knowledge transfer. Following this, this inaccurate information from jammed observations leads to suboptimal RB allocation decisions.
    \item \textbf{Policy Distortion}: The interference caused by the CJA and DRL-JA distorts the learning policy. Then the expert agent learns biased strategies due to the altered state-action feedback.
    \item \textbf{Increased Exploration-Exploitation Dilemma}: The CJA and DRL-JA affect the expert agent's learning to explore alternative actions more frequently. Then the agent struggles to distinguish between effective actions and those influenced by the jamming signal, impacting the exploitation of learned policies. 
    \item \textbf{Learning Instability}: Constant or inconstant interference can introduce instability in the expert agent's learning process. These changes in the system caused by the attacks hinder the convergence of the Q-learning algorithm.
    \item \textbf{Decreased Convergence Speed}: The presence of the jamming attack slows down the convergence speed of the Q-learning algorithm. Unreliable learning signals lead to longer training times and potentially prevent the achievement of an optimal resource allocation policy.
\end{itemize}

\subsection{Attack Mitigation Impact on CJA and DRL-JA}

As depicted in Fig. \ref{fig:Ex_Th_UE} to Fig. \ref{fig:Le_Le_HL}, the baseline mitigation method \cite{shi2021attack} properly mitigate the CJA, but it does not significantly reduce the effects of the DRL-JA on the system. It is critical to note that while the baseline mitigation method reduces the effects of the attacks, the system is unable to meet the requirements for UEs of the two slices. For this reason, an intelligent DRL-based mitigation strategy is proposed, which leaves the attacker confounded.
The DRL-based mitigation method, while perturbing the system, will not allow the attacker to interfere with the assigned RBs even if it perturbs the system. The proactive DRL-based mitigation method counters emerging threats in real-time, ensuring network slicing system robustness and reliability.

Fig. \ref{fig:Ex_Th_UE}, Fig. \ref{fig:Le_Th_UE}, Fig. \ref{fig:Ex_Le_UE}, and Fig. \ref{fig:Le_Le_UE} provide information about the effects of the traffic load on the system performance by using the baseline mitigation on these two attacks. As shown, the baseline mitigation method cannot reduce the effect of DRL-JA on the system, as preventing the system update does not stop the jammer from transmitting interference signals on the allocated RBs. If the jammer stops transmitting interference, the baseline method could use previous knowledge to reduce the attack's effect. However, as the jammer generates signals at varying intervals, the mitigation technique struggles to handle the attack's impact. Then the system requires a more sophisticated solution that changes its policy based on the attacker's policy. The intelligent solution mitigates the attack under various amounts of hidden layers and hidden units for the DRL-JA in Fig. \ref{fig:Ex_Th_HU} to Fig. \ref{fig:Le_Le_HL} depict the power of the mitigation method against various power of the jammer. DRL-JA has a more significant effect on the system and its effects on the system increase by having more hidden units and layers, and the baseline mitigation algorithm is not capable of reducing the effects of this attack. It, however, does not address the effects of the intelligent attack on the system; meanwhile, the intelligent solution can mislead the attacker and maintain the performance of the system. The DRL-based mitigation method misleads the jammer into generating signals on unallocated RBs. As a result, the SINR of allocated RBs remains unaffected even while the jammer transmits interference signals. However, if the jammer is able to predict the expert agent's RB allocation using DRL with deeper layers and more hidden units, the mitigation method would fail to mislead the jammer. For this reason, in \ref{subsection:FNN}, we consider the impact of both the jammer and the security administrator using feedforward neural networks (FNNs) \cite{shi2018adversarial}. It is important to note that these results are illustrated after the system converges. The model must be iterated extensively to reach convergence and learn the attacker's policy for generating signals.

\subsection{FNN and LSTM}
\label{subsection:FNN}

In this section, with the aim of more consideration about the effect on various neural networks, we consider the effect of using FNNs by the jammer and also for the system mitigation. We consider that the jammer uses an FNN, featuring two hidden layers, each consisting of 50 neurons for generating interference signal on the system. This network is trained using backpropagation with a cross-entropy loss function, and employs a softmax activation function at the output layer. The activation function for the hidden layers is sigmoid, and weights and biases are randomly initialized within the range of $[-1.0,1.0]$ \cite{shi2018adversarial}. The jammer utilizes the last 10 RB's allocation in the system for predicting the resource allocation of the system. Then it selects an RB for transmitting interference signals on it. The mitigation strategy by the FNN responds by altering the system and transmission signals to confuse the jammer and reduce its accuracy. For the mitigation side, the utilized FNN has two hidden layers, each comprising 60 neurons. Training is conducted via backpropagation with a cross-entropy loss function, and the output layer employs a softmax activation function. The hidden layers use a sigmoid activation function, while all weights and biases are initialized randomly within the range of $[-4.0,4.0]$. 

As listed in Table \ref{tab:attackersystem}, the choice between FNN and DRL for attackers significantly affects system performance. FNN, with its static architecture, struggles to capture temporal dependencies in resource allocation, making it less effective at predicting and disrupting RB allocations. While the proposed DRL-JA can generate interference signals on the allocated RBs and disrupt the system performance.

\begin{table}[ht]
    \centering
    \caption{Impact of Various NNs for Attackers on System Throughput and Latency}
    \label{tab:attackersystem}
    \small
    \begin{tabular}{|c|c|c|c|}\hline
       \multicolumn{4}{|c|}{FNN vs DRL on the DTRL System}\\\hline
        & \multicolumn{3}{c|}{System Degradation}\\\hline
       \multirow{3}{*}{Attacker NN} &  & Throughput & Latency  \\\cline{2-4}
        & FNN & 27\% & 36\% \\\cline{2-4}
        & DRL-JA & 54\% & 60\% \\\hline
    \end{tabular}
        \vspace{-0.3cm}
\end{table}

As listed in Table \ref{tab:nn} and Table \ref{tab:latency}, the type of NN used in the mitigation strategy also has effects on the system's performance. While the jammer utilizes FNNs, the security administrator can detect this behavior and understand that the jammer's signals rely only on the UEs' most recent requests. As a result, the mitigation system can effectively mislead the jammer. However, it should be noted that the model cannot predict all of the jammer's activities. Additionally, the jammer's generated signals increase interference in the system, leading to a reduction in performance compared to the no-attack scenarios.

\begin{table}[ht]
    \centering
    \caption{Impact of Various NNs for Attackers and Mitigation on the System Throughput}
    \label{tab:nn}
    \small
    \begin{tabular}{|c|p{1.3cm}|p{1.3cm}|p{1.3cm}|}\hline
       \multicolumn{4}{|c|}{FNN vs DRL on System Throughput}\\\hline
        & \multicolumn{3}{c|}{Mitigation NN}\\\hline
       \multirow{3}{*}{Attacker NN} &  & FNN & DRL  \\\cline{2-4}
        & FNN & 44\% & 91\% \\\cline{2-4}
        & DRL-JA & 27\% & 80\% \\\hline
    \end{tabular}
    \vspace{-0.3cm}
\end{table}

\begin{table}[ht]
    \centering
    \caption{Impact of Various NNs for Attackers and Mitigation on the System Latency}
    \label{tab:latency}
    \small
    \begin{tabular}{|c|p{1.3cm}|p{1.3cm}|p{1.3cm}|}\hline
       \multicolumn{4}{|c|}{FNNs vs RNNs on System Latency}\\\hline
        & \multicolumn{3}{c|}{Mitigation NN}\\\hline
       \multirow{3}{*}{Attacker NN} &  & FNN & DRL  \\\cline{2-4}
        & FNN & 41\% & 83\% \\\cline{2-4}
        & DRL-JA & 34\% & 70\% \\\hline
    \end{tabular}
\end{table}

In the DTRL system, the choice of NNs for both attackers and mitigation techniques greatly affects performance. When both attacker and mitigation use FNNs, the static nature of FNNs limits disruption and impact on latency and throughput. If the attacker uses DRL-JA while the mitigation uses FNN, the attacker’s advanced modeling increases disruption, leading to higher latency and reduced throughput. Using DRL-JA by the attacker and LSTM on the mitigation side results in a more dynamic interaction, with both systems adapting to temporal patterns, potentially causing a significant impact on system performance.

This paper studied the utilization of the DTRL algorithm which is optimized for allocating resources between eMBB and uRLLC slices. But introducing an additional slice like mMTC requires modifications to expand state and action spaces, and adjust reward functions for balanced QoS across all slices. Extending DRL-JA to multiple slices poses further challenges in monitoring and mitigating diverse jamming attacks across varying traffic types and resource allocations, demanding strategies for maintaining QoS under attack conditions.

\section{Conclusion}
\label{section:conclusion}

On the DTRL system, this paper examines the intersection of network slicing, jamming attacks, and mitigation strategies. By designing the smart jamming attack, this study examines the vulnerabilities of network slicing against jamming attacks. As a result of the intelligent DRL-based attack, the eMBB slice's throughput is reduced by 50\%, and the uRLLC slice's latency is increased by 60\%, for both expert and learner agents. As a result of this analysis and the proposed countermeasures, a vulnerability of network slicing is highlighted. To mitigate these attacks, an intelligent DRL-based method is proposed. By protecting the system against jamming-induced vulnerabilities, the mitigation method measures aim to enhance network slicing resilience, ensuring its ability to consistently deliver reliable services across a spectrum of applications and various data traffic. The method mitigates the attack and improves system performance by 80\% and 70\%, respectively. In the future, we will focus on the theoretical analyses of intelligent jamming attacks and mitigation techniques, aiming to theoretically analyze the performance of our proposed algorithms.

\section*{Acknowledgment}
This work has been supported by MITACS and Ericsson Canada, and the NSERC Canada Research Chairs program.

\bibliographystyle{unsrt}
\bibliography{references}
\end{document}